\documentstyle[prl,aps,multicol,epsfig]{revtex}
\begin{document}
\draft
\title{Wave Dynamical Chaos in a Superconducting Three-Dimensional 
       Sinai Billiard
       }
\author{H. Alt$^{1}$, C. Dembowski$^{1}$, H.-D. Gr\"af$^{1}$, 
        R. Hofferbert$^{1}$,\\
        H. Rehfeld$^{1}$, A. Richter$^{1}$, 
        R. Schuhmann$^{2}$, and T. Weiland$^{2}$\\
       }
\address{$^{1}$
         Institut f\"ur Kernphysik, Technische Hochschule Darmstadt,\\
         $^{2}$
         Fachgebiet Theorie Elektromagnetischer Felder, 
         Technische Hochschule Darmstadt,\\
         D-64289 Darmstadt, Germany\\
        }
\date{\today}
\maketitle
\begin{abstract}
Based on very accurate measurements performed on a superconducting
microwave resonator shaped like a desymmetrized 
three-dimensional (3D) Sinai billiard,
we investigate for the first time spectral properties 
of the vectorial Helmholtz,
i.e. non-quantum wave equation for a classically totally chaotic
and theoretically precisely studied system. 
We are thereby able to generalize some aspects of quantum chaos
and present some results which are consequences of the polarization
features of the electromagnetic waves.
\end{abstract} 
\begin{multicols}{2}
\narrowtext
For nearly 20 years, billiard systems have provided a very effective tool
for the investigation of semiclassical quantization of conservative
chaotic systems \cite {McDKau79,BerryBak}. 
This is due to the fact that even two-dimensional (2D)
billiards (as opposed to billiards of higher dimensionality)
are able to model a wide range of fully ergodic systems in
Gutzwiller's sense of ''hard chaos'' \cite {Gutzwiller}. 
As a matter of fact, properties of 
the wave dynamical spectra of such low-dimensional
but classically non-integrable systems are fully described 
by the Gaussian Orthogonal Ensemble (GOE) of Random Matrix Theory (RMT)
\cite {Mehta,Bohigas} if the underlying 
motion is invariant under time-reversal.
On the other hand, classically regular, i.e. integrable systems
lead to totally uncorrelated spectra.

Up to now investigations
on chaotic 3D-billiards were performed in experiments
with electromagnetic \cite {Deus,3Dcup} and acoustic
\cite {Weaver,ACME} waves, whereas the hardly feasible
numerical modelling was restricted to very special geometries
of high symmetry for the pure Schr\"odinger problem \cite {Harel}.
\begin{figure} [htb]
\centerline{\epsfxsize=8.6cm
\epsfbox{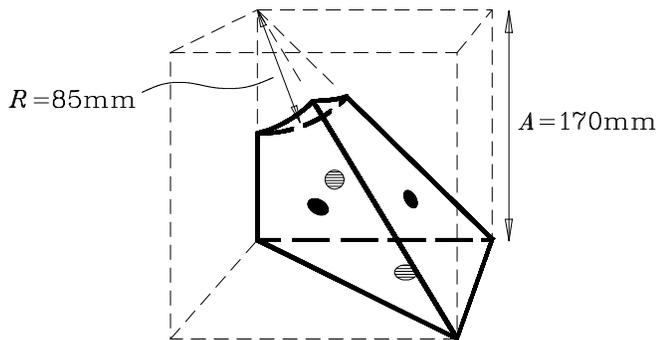}}
\caption{Geometry of the desymmetrized 3D-Sinai billiard
         (boldface line) which constitutes one sixth of the dashed cube.
         Eight of those cubes form the full system.
         As indicated in the figure, one antenna was located in the center
         of each plane surface of the cavity.
        }
\end{figure}

The goal of the present paper is to provide for the first time
a detailed analysis
of a fully chaotic three-dimensional (3D) 
electromagnetic billiard with a classically well-known
and theoretically precisely studied geometry: the 3D-Sinai billiard
resp its desymmetrized version given by 1/48 of a cube with a sphere
in its center, see Fig.1.
The system has to be described 
by the time-independent, fully vectorial Helmholtz equation
with electromagnetic boundary conditions.
The same geometry was recently investigated numerically in the quantum regime
\cite {Harel} described by the time-independent Schr\"odinger equation 
and experimentally with acoustic waves \cite {ACME}.
Our analysis therefore allows a very distinct comparison of 
totally different wave dynamical phenomena in a system with exactly the same
classical analogue. Our results should agree with those of \cite {Harel}
and \cite {ACME} if the conjecture holds that RMT is adequate to describe
spectra of arbitrary wave phenomena.

As several 2D- and 3D-billiards before \cite {3Dcup,Grapa,Hyperbel,widths},
the electromagnetic resonator was made 
of Niobium which becomes superconducting
below 9.2~K. This feature tremendously increases the resolution of the 
measured spectra due to quality factors of up to $10^7$ compared
to $10^3$ in normal conducting resonators and signal-to-noise
ratios of up to 70~dB. 
The resonator was cooled down to a temperature of
4.2~K 
and finally evacuated to a pressure of 
0.2~mbar inside a bath cryostat. 
We were able to excite the cavity via four different antennas up
to a frequency limit of 20~GHz given by our HP8510B vector network analyzer
which was used for the generation and detection
of the microwave signal.
Either the reflected signal from one antenna,
or the transmitted signal between two different antennas was measured.
Figure 2 shows a typical transmission spectrum between 6.50 and 6.75~GHz.
Identifying accurately per hand the positions of significant 
peaks of every single
spectrum and performing in the next step a detailed comparison of all the
possible spectra, a total set of 1881 experimental
eigenmodes was obtained which form
the base of the following investigations.

To check for the mechanical accuracy of the cavity and to estimate
the effects due to the coupling of the isolated resonator to the external
world (coupling holes and slightly penetrating coupling wires of the
rf-cables), we also simulated the spectrum using the electromagnetic
CAD-program MAFIA 
(solution of Maxwell's equation by the Finite Integration
Algorithm \cite {Weiland}). 
The resonator was modelled using a
mesh consisting of $10^6$ points, and a full 3D broadband time domain
calculation with a total number of 150,000 time steps was performed.
A task which is feasible on a standard workstation with one week of
CPU time.
The eigenfrequencies were finally determined by a Fast Fourier Transform
yielding a resolution of about 1.2~MHz over a wide frequency range.
Limited by this resolution
a comparison with the experimental spectrum (with a resolution of 10~kHz)
allows a one-to-one
correspondence between simulation and experiment within a region up to
7.5~GHz including 94 resonances.

For comparison, the numerically simulated eigenfrequencies are given
as dashed vertical lines in Fig.2. As a matter of fact the
experimental eigenfrequencies are systematically 
up-shifted due to the contraction of the 
resonator during the cool-down. 
From this shift (linearly increasing with frequency)
we extracted the relative contraction coefficient
$(\Delta x/x)_{exp}\approx 2.0\cdot 10^{-3}$ which is close to the
value for Niobium as given in the literature \cite{Erfling},
$(\Delta x/x)_{lit}\approx 1.4\cdot 10^{-3}$.
Beside this 
effect, every experimental resonance is individually shifted due
to the coupling holes and antennas and the mechanical fabrication tolerances.
However, these individual shifts are on the order of some
$5\cdot 10^{-4}$ and thus much smaller than the effect of the cool-down.

To prepare the experimental
spectrum of extracted eigenfrequencies for the statistical
analysis, we first unfolded it, i.e. rescaled the frequency axis
to a mean level spacing of unity. According to this, in applying the
generalized electromagnetic Weyl formula \cite {Lukosz,BalDup}
\begin{eqnarray}
N & & ^{smooth}(f) \approx \frac{8\pi}{3c_0^3}V\cdot f^3\label{3DWeyl}\\
& &-\bigg( \frac{4}{3\pi c_0}\int\frac{d\sigma}{R}-
\frac{1}{6\pi c_0}\int da\frac{(\pi-\Omega)
(\pi-5\Omega)}{\Omega}\bigg )\cdot f\nonumber\\
& & \nonumber\\
& & +const.\nonumber
\end{eqnarray}
to describe the smooth part of the integrated level density 
(spectral staircase), the 
spectrum is again checked for completeness. 
Here, $N$ counts the total number of resonances up to a certain
frequency $f$, $V$ denotes the volume of the geometry and $c_0$
is the speed of light. In the linear
term,  $R$ labels the
mean radius of curvature over the surface $\sigma$ and $\Omega$
is the dihedral angle along the edges $a$.
Note that the leading 
cubic term of this expression is twice the corresponding term 
in the scalar, i.e. quantum problem, and that
there is no quadratic term . Both features are due to the vectorial 
character of the electromagnetic field inside the cavity and thus
a consequence of two transverse polarizations of the electromagnetic wave. 
A comparision between Eq.(\ref{3DWeyl}) and the
integrated experimental level density 
yields an estimate for the number of uncertain resonances
$\Delta N\approx (V_{exp}/V_{theo}-1)
N_{total}\approx 4$. Here, $N_{total}$ denotes the complete set of 1881
resonances.
After the extraction of this smooth part
the expected non-systematic fluctuations around zero could be observed,
they are the carrier of all accessible information about the
classical dynamics inside the billiard.
\begin{figure} [hbt]
\centerline{\epsfxsize=8.6cm
\epsfbox{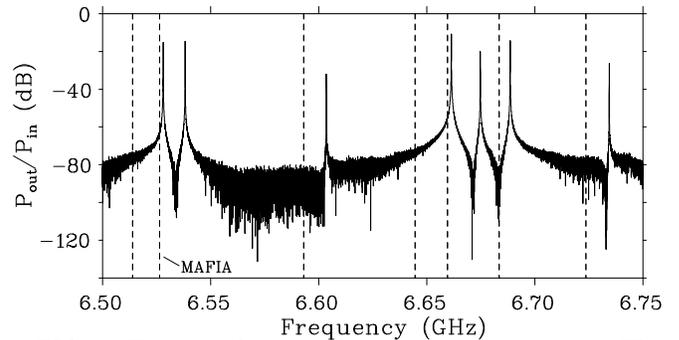}}
\caption{Excerpt of a typical transmission spectrum. The signal
         displays the ratio of output power to input power on a
         logarithmic scale. The numerically simulated
         eigenfrequencies are given by dashed lines. The shift between
         experiment and simulation is mainly
         due to the contraction of the resonator
         during the cool-down.
        }
\end{figure}

Statistical measures developed to display the correlations embedded
in these fluctuations were originally introduced in the 1950th
to describe spectra of nuclear systems with many degrees of freedom.
\cite {Mehta}. 
An adequate
and widely tested statistical measure to examine short-range correlations
up to a length of {\it one} mean level spacing is given by the
nearest neighbour spacing distribution (NND), displaying the
probability $P(s)$ for a certain spacing $s$ of two adjacent resonances
in the spectrum. Figure 3 (l.h.s.) 
displays $P(s)$ of the experimental spectrum
as well as the two limiting curves, ''Poisson'' for a totally
uncorrelated spectrum corresponding to regular motion on the classical side,
and ''GOE'' for the totally chaotic case. Obviously, the experimental
histogram is very close to the GOE distribution. 
In the next step we analyzed the spectrum on a larger 
scale in order to investigate long-range correlations. For this
purpose we calculated $\Sigma^2(L)$, which expresses the variance
of a number of resonances inside an interval of length L on the
unfolded scale, as well as the related Dyson-Mehta statistics,
$\Delta_3(L)$, also sensitive up to L, i.e. 
{\it several} mean level spacings. The result for both properties is
given as well in Fig.3 (l.h.s.). Here, two observations can be made: First,
the experimental curves rapidly deviate from the GOE prediction
and lie in between the regular and the chaotic case, and second,
above a certain value $L_{max}$, which is different for both
statistics ($L_{max}^{\Sigma^2}\approx 40,~L_{max}^{\Delta_3}\approx 150$), 
the experimental curves run into saturation. This last feature
is exactly what is expected from theory \cite {Berry400}, displaying
the fact that for increasing $L$ the given statistics are more and more
sensitive to specific, i.e. non-universal features of the system.
In this sense $L_{max}$ separates the universal scale
which is dominated by the general dynamical behaviour of the classical
analogue -- described by the collective behaviour of all {\it long}
periodic orbits of the system -- from the non-universal scale which is
generated by individual system-specific attributes -- in the present
case given by the {\it shortest} periodic orbits of the billiard. 
Note that the difference between $L_{max}$ of $\Sigma^2$ 
and the value obtained for $\Delta_3$ is 
$L_{max}^{\Delta_3}\approx 4\cdot L_{max}^{\Sigma^2}$ \cite {Delon},
which is well reproduced by the given spectrum. 
Theoretically the length of
the shortest periodic orbits $l_{min}$ depends on $L_{max}^{\Delta_3}$
according to \cite {3Dcup}
\begin{equation}
L_{max}^{\Delta_3}=\frac{3c_0}{l_{min}f_{max}}\frac{N_{total}}{2}~,
\end{equation}
where $f_{max}$ denotes the upper frequency limit of 20~GHz.
Using the length of the first observable periodic orbit, $l_{min}$=0.34~m 
(see below),
one estimates $L_{max}^{\Delta_3}\approx 125$, which is 
close to the value of Fig.3.
\begin{figure} [hbt]
\centerline{\epsfxsize=8.6cm
\epsfbox{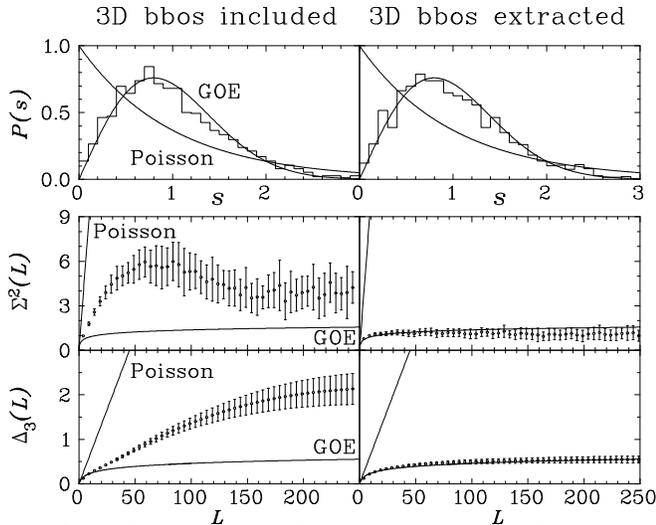}}
\caption{Short-range and long-range statistical measures before and
         after the extraction of 3D bbos, see text.
        }
\end{figure}
To characterize all statistical measures more quantitatively and
in order to check {\it how well} the GOE prediction describes the results,
we also analyzed the experimental curves according to a model of Berry
and Robnik \cite {BerRob} within the universal region. 
The basic intention of this model is to provide
a continuous interpolation between the pure Poissonian and GOE using
a mixing-parameter $q$ which corresponds to the relative phase space volume
that is covered by chaotic trajectories. In the present case we obtained
for the three statistical measures on the l.h.s. of Fig.3
$q_{NND}=0.95\pm 0.01,~q_{\Sigma^2}=0.89\pm 0.02$ and 
$q_{\Delta_3}=0.91\pm 0.02$,
indicating a small but systematic deviation from unity, the limit for
pure chaotic systems. Thus, although the system is fully ergodic on the
classical side without any stable islands in phase space,
the wave dynamical side pretends their existence. To understand this
phenomenon, we analyzed the classical analogue in more detail.
Therefore, we continued our investigation on a more specific
scale which provides a bridge 
between the classical chaotic features and their impact on
the electromagnetic spectrum. 
This scale is given by the length spectrum of the billiard, which
is practically obtained through the Fourier
transform of the spectral level density or its fluctuating part, respectively,
\begin{equation}
\tilde{\rho}^{fluc}(l)= \int_{f_{min}}^{f_{max}}\rho^{fluc}(f)
\cdot \exp({i\frac{2\pi}{c_0}lf})\,df~.
\label{Fourint}
\end{equation}
Here $l$ denotes the physical length scale. Since the property 
$\tilde{\rho}^{fluc}(l)$ maps the long periods in frequency space
onto short scales in length space, the resulting spectrum shows peaks near
the classical periodic orbits of the billiard. Figure 4 exhibits the lower
part of this length spectrum up to $l$=1.5~m. Here a rich structure
of peaks can be observed above a minimum length $l_{min}$=0.34~m. This
first peak belongs to the shortest 3D bouncing ball orbit (3D bbo) of the
billiard, propagating along one edge of the desymmetrized cube without
striking the sphere, see sketch in the inset. 
\begin{figure} [hbt]
\centerline{\epsfxsize=8.6cm
\epsfbox{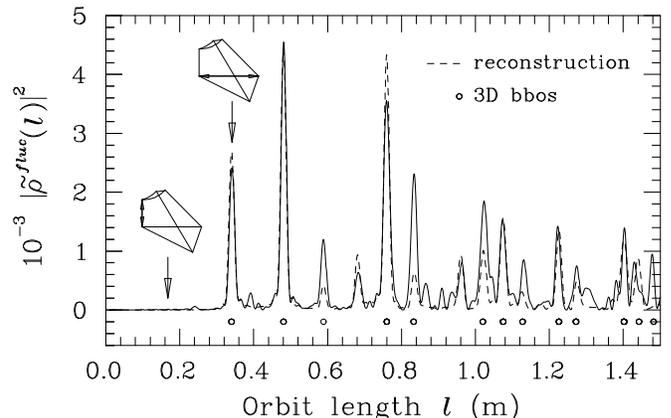}}
\caption{Experimental length spectrum of the billiard (full line) 
         and the semiclassical reconstruction using only 3D bbos
         (dashed line). The pictures in the inset show the first
         unstable and the first 3D bouncing ball orbit, respectively. 
        }
\end{figure}
Exactly this shortest length was used 
to estimate the saturation parameter $L_{max}$ of $\Delta_3$
(see above). In fact, this first 3D bbo is the shortest non-isolated,
neutrally stable periodic orbit of the system. It is highly remarkable,
that this is not the first periodic orbit of the billiard.
The shortest unstable periodic orbit runs along the shortest edge
of the desymmetrized billiard which is intersected by the sphere, it
possesses a length $l_{min}^{upo}$=0.17~m, see sketch in the inset. 
The corresponding peak
in the Fourier spectrum is drastically suppressed due to the desymmetrization
of the system itself \cite {Harel,Eckhardt}. 
As a matter of fact, the obtained length spectrum is totally dominated
by bbos of all possible dimensions, not only in the quantum case
\cite {Harel} but also in the electromagnetic counterpart.
To demonstrate this, we considered the contribution
of the leading 3D bbos to the given length spectrum. Therefore, we 
used a lattice vector description of Berry \cite {BerrySinai} 
to label and generate all
3D bbos up to a certain length and determined their contribution 
on $\rho^{fluc}_{bbo}(f)=dN^{fluc}_{bbo}/df$ through \cite {3Dcup}
\begin{equation}
{N}^{\rm fluc}_{bbo} (X) = \frac{2\pi S_{bbo}}{l_{bbo}^2}\bigg(
\sum_{0 < n < X}
\big(X^2 - n^2 \big) - \frac{2}{3} X^3 +\frac{1}{2} X^2 \bigg)~,
\label{3Dbbo}
\end{equation}
with $X=l_{bbo}f/c_0$. Here, the length of a given 3D bbo, $l_{bbo}$, 
was deduced directly from the lattice vector and $S_{bbo}$, the
perpendicular area on which this orbit exists, was fixed in a 
Monte-Carlo simulation.
In the given range up to $l$=1.5~m we obtained 55 
3D bbos of different degeneracies and with positive $S_{bbo}$,
their superimposed semiclassical reconstruction due to Eq.(\ref{3Dbbo})
is given by the dashed line in Fig.4. It is highly
remarkable that nearly the full structure of the given length
spectrum can be reproduced using only 3D bbos, whereas the
influence of the enormous number of unstable periodic orbits
(approximately 36000 up to $l$=1.5~m \cite {Harelpriv})
is hidden in the background. Discrepancies between the experimental
length spectrum and the reconstruction can be 
predominantly found at the locations 
of the 3D bbos themselves and arise because
of the existence of sub-dimensional and tangential bbo manifolds 
\cite {Harel}. 

Finally, to demonstrate the influence of the considered 3D bbos on the 
statistical measures NND, $\Sigma^2$ and $\Delta_3$, we repeated our 
statistical analysis using a modified unfolding procedure
in which the standard Weylian, Eq.(\ref{3DWeyl}), is extracted
together with the contribution of the 3D bbos, Eq.(\ref{3Dbbo}).
For this we used all 3D bbos up to a length $l$=3.0~m.
The result is given in Fig.3 (r.h.s.) displaying, 
now also for $\Sigma^2$ and $\Delta_3$, 
nearly perfect agreement with the GOE prediction in the
universal regime up to $L_{max}$. The corresponding, i.e.
corrected mixing parameters, were fixed to be  
$q_{NND}=0.96\pm 0.01,~q_{\Sigma^2}=0.99\pm 0.01$ and 
$q_{\Delta_3}=0.99\pm 0.01$.

In summary, a set of 1881 highly resolved eigenmodes
of an electromagnetic 3D-Sinai billiard was analyzed according to
standard methods of Random Matrix and Periodic Orbit Theory.
Spectral correlations are shown to be totally consistent
with the predictions of the GOE after the
systematic extraction of the family of 3D bbos which
dominate not only the length spectrum of the billiard but 
also lead to dramatic deviations from Gaussian characteristics
in the measures for long-range correlations, $\Sigma^2$ and $\Delta_3$.

We would like to thank H. Lengeler and the CERN workshops for the
excellent fabrication of the cavity. We are 
very grateful to B. Eckhardt and H. Primack for a lot of suggestions,
discussions and their theoretical cooperation. One of us (A.R.) thanks
in particular U. Smilansky for providing him first with a 
one-to-one paper model of the desymmetrized 3D Sinai billiard of
Fig.1 which led to the present investigations. We finally 
thank M. K\"oppen for providing us with the thermal contraction coefficient
of niobium. 
This work has been supported by the Sonderforschungsbereich 185
``Nichtlineare Dynamik'' of the Deutsche Forschungsgemeinschaft and in
part by the Bundesministerium f\"ur Bildung und Forschung under
contract number 06DA665I.

\end{multicols}

\begin{references}
\bibitem{McDKau79}   S.W. McDonald and A.N. Kaufman, 
                     Phys. Rev. Lett. {\bf 42}, 1189 (1979).
\bibitem {BerryBak}  M.V. Berry, 
                     Proc. R. Soc. Lond. A{\bf 413}, 183 (1987).
\bibitem{Gutzwiller} M.C. Gutzwiller, 
                     {\it Chaos in Classical and Quantum
                     Mechanics} (Springer, New York, 1990).
\bibitem{Mehta}      M.L. Mehta, 
                     {\it Random Matrices}, 2nd ed.,
                     (Academic Press, San Diego, 1991).
\bibitem{Bohigas}    O. Bohigas, 
                     in {\it Chaos and Quantum Physics}, 
                     M.-J. Giannoni, A. Voros, and J. Zinn-Justin, eds.
                     (Elsevier, Amsterdam, 1991), p. 89.
\bibitem{Deus}       S. Deus, P.M. Koch and L. Sirko, 
                     Phys. Rev. E{\bf 52}, 1146 (1995).
\bibitem{3Dcup}      H. Alt, H.-D. Gr\"af, R. Hofferbert,
                     C. Rangacharyulu, H. Rehfeld, A. Richter, 
                     P. Schardt, and A. Wirzba,
                     Phys. Rev. E{\bf 54}, 2303 (1996).
\bibitem{Weaver}     R.L. Weaver, 
                     J. Acoust. Soc. Am. {\bf 85}, 1005 (1989).
\bibitem{ACME}       C. Ellegaard, T. Guhr, K. Lindemann, H.Q. Lorensen,
                     J. Nyg\aa rd, and M. Oxborrow,
                     Phys. Rev. Lett. {\bf 75}, 1546 (1995).
\bibitem{Harel}      H. Primack and U. Smilansky,
                     Phys. Rev Lett. {\bf 74}, 4831 (1995).
\bibitem{Grapa}      H.-D. Gr\"af, H.L. Harney, H. Lengeler, 
                     C.H. Lewenkopf, C. Rangacharyulu, A. Richter, 
                     P. Schardt, and H.A. Weidenm\"uller,
                     Phys. Rev. Lett. {\bf 69}, 1296 (1992).
\bibitem{Hyperbel}   H. Alt, H.-D. Gr\"af, H.L. Harney, R. Hofferbert,
                     H. Lengeler, C. Rangacharyulu, A. Richter, 
                     and P. Schardt,
                     Phys. Rev. E{\bf 50}, 1 (1994).
\bibitem{widths}     H. Alt, H.-D. Gr\"af, H.L. Harney, R. Hofferbert, 
                     H. Lengeler, A. Richter, P. Schardt,
                     and H.A. Weidenm\"uller,
                     Phys. Rev. Lett. {\bf 74}, 62 (1995).
\bibitem{Weiland}    T. Weiland, 
                     {\it International Journal of Numerical Modelling},
                     {\bf 9}, 295 (1996).
\bibitem{Erfling}    H.D. Erfling, 
                     Ann. Physik {\bf 41}, 467 (1942).
\bibitem{Lukosz}     W. Lukosz,
                     Z. {\it Physik} {\bf 262}, 327 (1973).
\bibitem{BalDup}     R. Balian and B. Duplantier, 
                     Ann. of Phys. {\bf 104}, 300 (1977).
\bibitem{Berry400}   M.V. Berry, 
                     Proc. R. Soc. Lond. A{\bf 400}, 229 (1985).
\bibitem{Delon}      A. Delon, R. Jost and M. Lombardi,
                     J. Chem. Phys. {\bf 95}, 5701 (1991).
\bibitem{BerRob}     M.V. Berry and M. Robnik, 
                     J. Phys. A{\bf 17}, 2413 (1984).
\bibitem{Eckhardt}   O. Frank and B. Eckhardt, 
                     Phys. Rev. E{\bf 53}, 4166 (1996).
\bibitem{BerrySinai} M.V. Berry, 
                     Ann. Phys. {\bf 131}, 163 (1981).
\bibitem{Harelpriv}  H. Primack, 
                     private communication (1996).
\end{references}
\end{document}